# Gamma-rays from the funnels of thick accretion disks around active galactic nuclei

W. Bednarek* and J. G. Kirk

Max-Planck-Institut für Kernphysik
Postfach 10 39 80
D-69029 Heidelberg
Germany

**Abstract.** We analyse the propagation of particles in the narrow funnel of a thick accretion disk. It is assumed that: (1) the funnel walls emit black body radiation with temperature decreasing outwards; (2) the magnetic and electric fields are longitudinal in the funnel. Such a scenario is consistent with models in which a large potential drop is induced by a rotating massive black hole threaded by a magnetic field (Blandford & Znajek 1977, Macdonald & Thorne 1982). The interaction of relativistic protons with thermal photons from the funnel results in direct production of $e^{\pm}$ pairs and/or pions. We discuss the relative importance of these processes for different conditions in the funnel (temperature and electric field profiles). Injected $e^{\pm}$ pairs interact with thermal photons in the Thomson or Klein-Nishina regime. Under some conditions, a pair avalanche results, which we assume saturates in the production of stable bunches, containing almost equal numbers of electrons and positrons. As a consequence, highly collimated $\gamma$-ray photons are produced. We obtain $\gamma$-ray spectra from our model in order to test its applicability to the $\gamma$-ray emitting AGNs recently reported by EGRET.

**Key words:** Gamma rays: theory – Galaxies: active: individual: CTA 102, QSO 1633+382 (4C 38.41)

## 1. Introduction

The recent discovery of high energy $\gamma$-ray emission ($> 100$ MeV) from the class of active galactic nuclei (AGNs), known as 'blazars', has demonstrated unambiguously that high energy processes play a fundamental role in AGNs. Other types of AGN, which have previously been reported as sources of soft $\gamma$-rays (e.g. the Seyfert galaxies: MGC8-11-11, NGC 4151) have not been detected above 1 MeV by the SIGMA telescope or the OSSE and COMPTEL telescopes of the Compton Gamma Ray Observatory (Jourdain et al. 1992; Maisack et al. 1993), which can be interpreted as evidence for the variability of the high energy emission of these sources.

*Send offprint requests to*: J. G. Kirk
*present address: University of Łódź, Łódź, Poland

Models of $\gamma$-ray production in AGNs can be characterised by the geometry they assume for the central region (spherically symmetric or jet-like being the most popular choices) and by the basic mechanism of $\gamma$-ray production: hadronic or leptonic. The older hadronic models postulate the acceleration of protons (usually shock acceleration is suggested) up to extremely high energies in an isotropic picture. The interaction of protons with matter and/or radiation then leads to the production of $e^{\pm}$ pairs (e.g., Sikora et al. 1987), neutrons (e.g., Kirk & Mastichiadis 1989; Sikora et al. 1989) and pions (e.g., Morrison et al. 1984; Bednarek & Calvani 1991). Energetic secondary $e^{\pm}$ pairs and neutrons subsequently participate in the generation of $\gamma$-rays. However, in the case of blazars, the discovery of significant variability on a time scale of days, coupled with extremely high fluxes of $\gamma$-rays leads to severe problems with the energy budget in isotropic models, which require luminosities $\sim 10^{48}\,\mathrm{erg\,s^{-1}}$. Following the work of Marscher (1980), Königl (1981), Reynolds (1982), Ghisellini et al (1985) and Ghisellini & Maraschi (1985), Maraschi et al (1992) and Zdziarski & Krolik (1993) have suggested anisotropic models in which the emission results from the synchrotron self-Compton mechanism operating in electron blobs moving with relativistic speed. Other similar models suggest the comptonisation by such a blob of soft radiation from the accretion disk (e.g., Melia & Königl 1989; Dermer et al. 1992, Dermer & Schlickeiser 1993; Sikora et al. 1993). Recently, anisotropic hadronic models have also been proposed in which the $\gamma$-rays originate in the inner jet as a result of relativistic protons colliding with soft photons of synchrotron origin (Mannheim et al 1991, Mannheim & Biermann 1992) or with matter of a thick accretion disk (Bednarek 1993).

In this paper we examine a new anisotropic picture for the high energy phenomena observed in AGNs, in which the interaction and acceleration of electrons/positrons occurs deep in the narrow funnel of a thick accretion disk surrounding the central black hole. The inner parts of the funnel contain little matter, because the magnetic field lines there thread the event horizon of the black hole and effectively exclude material from the accretion disk. However, the region is permeated by radiation emitted by the walls. This we expect to be of approximately Planckian spectrum and to have a temperature near the base of the funnel significantly higher than that indi-cated by the observations of the UV bump, but consistent with observations of soft X-ray excesses in some AGNs. The main difference between the scenario investigated here and previous work is that we assume charged particles to be accelerated by a component of the electric field parallel to the magnetic field. Such a configuration is possible only where the assumption used in ideal MHD of infinite conductivity breaks down, which can occur locally at sites of magnetic reconnection, or, globally, if insufficient charge is available to screen out the electric field induced by a rotating magnetic field. Most of our calculations apply equally well to each of these cases. But, where necessary, we concentrate in this paper on the latter picture and assume that both the electric and magnetic fields lie along the axis of the funnel, which contains no background plasma. Henri and Pelletier (1991) have developed a model using a similar overall picture and propose the outer parts of the funnel contain a subrelativistic jet. However, in contrast to our model, they assume the inner core of the funnel contains a pair plasma of high enough density to short out the induced electric field.

The acceleration of protons by the electric field in our scenario is limited either by energy losses in direct $e^{\pm}$ pair production or by losses in pion production on photons, depending on the parameters in the funnel. The acceleration of electrons or positrons, on the other hand, is limited by inverse Compton losses which can efficiently convert energy extracted from the electric field into high energy radiation. Only for very strong electric fields and low radiation temperatures does the process of triplet pair production take over as the major energy loss mechanism. Synchrotron radiation, which would play a role if particles were injected isotropically, is strongly suppressed because pitch angles remain small ($\sim 1/\gamma$) throughout the cascade. The details of the physical scenario and the description of the interaction of protons and electrons with thermal photons are discussed in Section 2. We find that the comptonisation of thermal radiation by electrons in the Klein-Nishina regime results in the formation of bunches of pair plasma. Such bunches move through the funnel and scatter thermal photons into the high energy $\gamma$-ray range in a way reminiscent of the pulsar model described by Michel (1991). In terms of the bunch scenario, the $\gamma$-ray emission from two blazars the first with a hard and the second with a soft $\gamma$-ray spectrum (QSO 1633+382 and CTA 102, respectively), is modelled in Section 3. Finally, in Section 4 we summarise and discuss our results.

## 2. High energy radiation from the funnel of an accretion disk

We start from a picture of the central region of an AGN in which a geometrically thick, radiation supported accretion disk exists around a supermassive black hole (for a review see Begelman et al. 1984). Studies of the structure of such a disk postulate the existence of a narrow funnel on the rotation axis. The funnel is filled with thermal radiation (we assume a black body distribution) which is emitted by the disk walls (e.g., Jaroszyński et al. 1980, Abramowicz et al. 1980, Madau 1988). This picture of the central region of an AGN is consistent with observations of a UV bump in the spectrum of some sources (Shields 1978; Malkan & Sargent 1982). In addition, there is observational evidence of a soft X-ray excess in some AGNs (e.g., Yaqoob & Warwick 1991; Puchnarewicz et al. 1992 and references therein). A possible interpretation of this excess is that it comes from the inner part of the funnel where the char-acteristic temperature is significantly higher than that of the region which gives rise to the UV bump. In some AGNs – in particular in blazars – this component may be reduced because of the narrow opening angle of the funnel; also, it may simply be too difficult to detect this component in AGNs at high redshifts, because it will be shifted to lower photon energies and absorbed en route. On the other hand, some sources which show no evidence of beaming do have a soft X-ray excess. Clearly, observations are inconsistent with the hypothesis of a unique opening angle for the funnels of all AGN. Nevertheless, a significant temperature gradient is indicated in the disk funnel, and we incorporate this into our model. A similar approach has been taken by Madau (1988), who adopts a temperature profile ranging between a few times $10^6$ K in the inner regions and $\sim 10^4$ K in outer regions of the funnel in order to fit the UV – soft X-ray excess seen in the spectra of some AGNs.

Inside the funnel, we assume that particles are accelerated in a constant electric field. Large potential drops can be induced by a supermassive black hole threaded by a magnetic field predominantly parallel to its rotational axis (Blandford & Znajek 1977; Macdonald & Thorne 1982, for reviews see Begelman et al. 1984; Rees 1984). This field, which is maintained by currents flowing in the funnel walls can be quite strong (of the order $10^4$ G) and prevents plasma from the disk or funnel walls from penetrating into the inner parts of the funnel where the magnetic field lines are linked to the horizon of the black hole. As a result, this region of the funnel may possess insufficient charge to screen out the induced electric field. If only small amount of *primary* plasma is injected into this region of the funnel, (e.g., $e^{\pm}$ pairs from $\gamma$-$\gamma$ annihilation or e-p plasma from the decay of low energy neutrons) the injected particles will be accelerated by the electric field.

A large scale electric field can also be induced by an accretion disk rotating in a perpendicular magnetic field (see e.g., Lovelace 1976; Blandford 1976). However, in this case the magnetic field lines pass through the inner disk, so that plasma is able to enter and may screen out the induced electric field. Alternatively, it is also possible that smaller scale electric fields are formed inside a disk or a jet, at points where magnetic reconnection takes place (e.g., Biskamp 1989). Although it is difficult to make a quantitative calculation, estimates from one recent model suggest electric fields of the order of $3.10^8$ V cm$^{-1}$ (Haswell et al 1992). The discussion below of the equilibrium Lorentz factor achieved in an electric field applies to this picture too, since the minimum size required in order for the pair production avalanche to be important is extremely small. However, the degree of collimation of the accelerated particles in the reconnection picture is uncertain, and we base the calculations in this paper on the case of a large scale field aligned along the funnel axis.

We consider, then, the propagation of particles in the narrow funnel of an accretion disk with longitudinal electric field $E^f$ (measured in V cm$^{-1}$) and longitudinal magnetic field. The orientation of the vectors of angular momentum of the disk and magnetic field in the funnel is chosen to be antiparallel, allowing the acceleration of particles with positive charge outwards, although this is important only for proton acceleration. We suppose that there exists a mechanism of particle injection into the electric field deep in the funnel and discuss the fate of injected protons and leptons in turn.

## 2.1. Injection of protons

The important interactions of protons with soft photons are direct production of $e^\pm$ pairs ($p + \gamma \to e^\pm$) and photo-pion-production ($p + \gamma \to \pi$). Figure 1 shows the energy loss rates $P(p\gamma \to \pi/e^\pm)$ of a proton in a black body radiation field by each of these processes for three different radiation temperatures, computed from formulae given by Blumenthal (1970) and Stecker (1968). In the case of pair production, the approximation is made that most interactions occur near threshold, i.e., the energy of the incoming photon is neglected. A more accurate treatment has been given by Begelman et al (1990), but the differences (up to a factor of 2 for the highest temperature radiation field) are unimportant for our analysis.

At any given temperature, there exists a value of the proton energy above which losses by pion production are dominant and below which the losses are determined by pair production. The transition point between these two processes corresponds to a specific energy loss rate and, consequently, to a value of the electric field strength in the funnel. In stronger fields protons lose energy mainly by pion production in weaker ones by pair production. The dependence of this critical value of the electric field $E^f_{\pi/e^\pm}$ in the funnel as a function of the radiation temperature is given by,

$$E^f_{\pi/e^\pm} \approx 2.5 \times 10^{-8} \cdot T^2 \ [\mathrm{V\,cm^{-1}}] \quad (1)$$

In Fig. 2 this relation is represented by the solid line. The Lorentz factor of a proton at which gains are balanced by losses determines the typical energy at which pairs are produced. In the region in which pion production dominates, pairs are produced by the hard gamma-rays resulting from $\pi^0$-decay when they interact with a photon of the blackbody radiation field. The Lorentz factor of such pairs is about 100 times the proton Lorentz factor and they decelerate not by inverse Compton scattering (ICS), but primarily by producing more pairs in the triplet pair production process (Mastichiadis 1991), thus forming an electromagnetic cascade in which few hard photons are present. This cascade continues until a bunch is formed (see below), since the ICS process is in the Klein-Nishina regime for funnel parameters appropriate to losses by pion production (see Fig. 2). One interesting consequence of the process of pion production is the conversion of the relativistic protons into relativistic neutrons, which move ballistically and gain no further energy from the electric field. After only a few pion producing interactions, essentially all protons will have been converted into neutrons. If, as we assume, the radiation temperature in the funnel decreases outwards, the neutrons will not interact with thermal radiation in their path, but escape freely from the funnel, most of them decaying outside the central core of the AGN at a distance determined by their Lorentz factors ($R_n[\mathrm{cm}] \approx 3 \times 10^{13} \cdot \gamma_n$).

If the interaction of protons occurs deep in the funnel where $T$ and $E^f$ are below the line $p\gamma \to \pi/e^\pm$, then $e^\pm$ pairs are produced directly with Lorentz factors comparable to those of the primary protons. However, in contrast to the case of pion production, protons cannot convert to neutral particles and escape, but proceed to inject pairs throughout the whole funnel. The secondary pairs are decelerated in the funnel by Compton scattering until they achieve equilibrium either as single particles at $\gamma_{eq}$ (for parameters below the line KN-T in Fig. 2) or as members of a bunch.

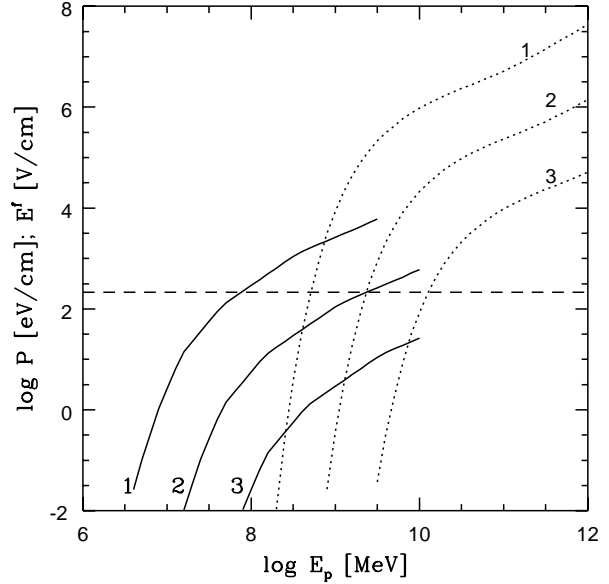

**Fig. 1.** The energy loss of protons $P[\mathrm{eV/cm}]$ per unit distance travelled in the funnel as a function of proton energy. (At saturation we have $P = E^f$.) Shown are losses by pion production (dashed line) and $e^\pm$ pair production (solid line) in a black body radiation field with temperatures $T = 10^{5.5}$K (1), $T = 10^5$K (2) and $T = 10^{4.5}$K (3). The horizontal dashed line defines the value of electric field $E^f$ (for $T = 10^5$K) below which protons lose energy mainly by $e^\pm$ pair production and above which pion production dominates the losses

## 2.2. Injection of electrons and positrons

Let us assume that positrons of relatively low energy are injected into the electric field in the funnel. They are accelerated by the component of the electric field parallel to the magnetic field until, at an energy $\gamma_{eq} m_e c^2$ which depends on the local temperature of the radiation field $T$ and the electric field strength $E^f$, their rate of energy gain is balanced by energy losses due to inverse Compton scattering. Synchrotron radiation, which can compete with inverse Compton losses for isotropically distributed particles if $B \gtrsim 4.10^{-7} T^2$ is unimportant for the accelerated particles, since these conserve the component of their momentum perpendicular to the magnetic field during acceleration. On scattering, the pitch angle is increased to at most a value $\sim 1/\gamma$.

If the scattering takes place in the Thomson regime, i.e., if $\gamma_{eq}(3k_B T/m_e c^2) \leq 1$, the equilibrium Lorentz factor ($\gamma_{eq}$) is given by

$$\gamma_{eq} \approx 1.5 \times 10^{13} \cdot \left(E^f\right)^{1/2} T^{-2}. \quad (2)$$

In the conditions envisaged, positrons attain this Lorentz factor essentially instantaneously, after travelling a distance

$$d_{eq} \approx 10^{19} (E^f)^{-1/2} T^{-2} \ \mathrm{cm}, \quad (3)$$

which is much smaller than any macroscopic dimension of interest.

Positrons coasting with Lorentz factor $\gamma_{eq}$ in the Thomson regime transfer energy gained from the electric field directly into high energy photons by ICS. However, in some parts of the funnel it is possible for the scattering to occur in the Klein-Nishina regime where $\gamma_{eq}(3k_B T/m_e c^2) \geq 1$. The line dividing these two parts of the parameter space $(T, E^f)$ is depicted in Fig. 2 and was obtained by setting the average energy of thermal photons in the reference frame of the electron (i.e., $3k_B T\gamma_{eq}$) equal to the electron rest-mass energy. This gives the critical electric field $E^f_{KN-T}$ as a function of temperature:

$$E^f_{KN-T}(T) \approx 1.6 \times 10^{-8} T^2. \quad (4)$$

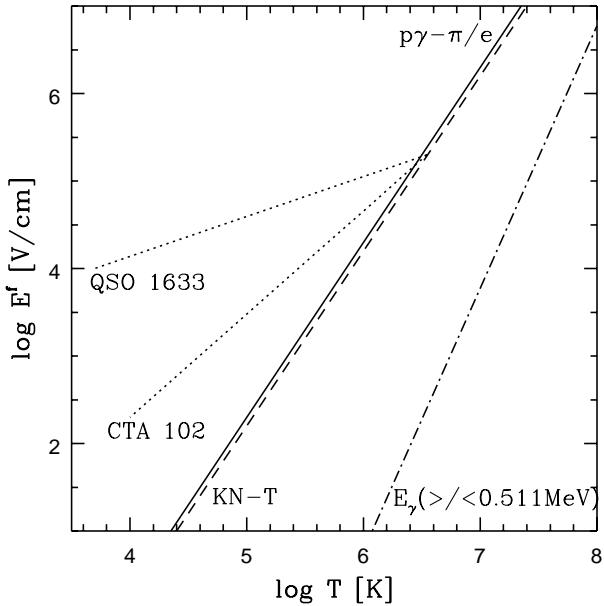

Fig. 2. The separation of the different modes of interaction of protons and electrons in the narrow funnel of a thick accretion disk. The parameter space is defined by the local values of the electric field in the funnel $E^f$ and the temperature of radiation $T$. The solid line $p\gamma - \pi/e$ separates regions in which protons lose energy mainly by pion production (above this line) and $e^{\pm}$ pair production (below this line). The dashed line KN-T separates the regions in which single electrons move in the Klein-Nishina regime (above) and Thomson regime (below). Above the dash-dotted line $E_\gamma(> / < 0.511 \text{ MeV})$ electrons comptonise thermal photons to energies $> 0.511$ MeV (these photons may produce additional $e^{\pm}$ pairs in $\gamma - \gamma$ collisions). The two dotted lines give the range of parameters used in the models of CTA 102 and QSO 1633+382

It is also interesting to find out for which parameters $(T, E^f)$ most of the photons produced in the comptonisation process have energies above 0.511 MeV. If such photons are created by both electrons and positrons, they are beamed in opposite directions, and are capable of producing additional $e^{\pm}$ pairs. We can estimate this condition easily by applying the formula for the average energy $\langle E_\gamma \rangle$ of scattered photons:

$$\langle E_\gamma \rangle = \frac{4}{3}(3k_B T)\gamma_{eq}^2, \quad (5)$$

Replacing $\gamma_{eq}$ in Eq. (5) by the value from Eq. (2), we find for the critical field $E^f_{511}(T)$ the relation

$$E^f_{511}(T) \approx 6.2 \times 10^{-18} T^3. \quad (6)$$

The corresponding critical line in parameter space is indicated in Fig. 2 by the dot-dashed line.

An examination of Fig. 2 gives us an idea about the conditions in the funnel required for efficient production of $\gamma$-ray photons from inverse Compton scattering of low energy thermal photons by relativistic positrons. If electrons and positrons are injected at a location deep in the funnel where $\gamma_{eq}$ lies in the Thomson regime, the $\gamma$-ray spectrum can be obtained simply by integrating the contributions of positrons (the outgoing component) moving with Lorentz factor equal to the local value of $\gamma_{eq}$ throughout the funnel, provided the probability of interaction between two $\gamma$-rays is sufficiently small. This restriction depends on both the luminosity of the source and the geometry of the funnel. For the case of blazars, we can make a simple estimate by taking a funnel of length $r_{max} = 10^{16}$ cm and allowing for a flux of photons corresponding to an isotropic luminosity of $L = 10^{48}$ erg s$^{-1}$ in gamma-rays of energy $E_\gamma \geq 100$ MeV. The photon number density in the funnel is then

$$n_\gamma \approx \frac{L}{4\pi r_{max}^2 c E_\gamma} \quad (7)$$

$$\approx 2.10^8 \text{ cm}^{-3} \quad (8)$$

If we assume equal numbers of these photons travel inwards and outwards, the optical depth through the funnel to pair production in a head-on $\gamma$-$\gamma$ collision is roughly

$$\tau_{\gamma\gamma} \approx n_\gamma \sigma_{\gamma\gamma} r_{max} \quad (9)$$

$$\lesssim 2 \times 10^{-4} \quad (10)$$

where we have taken the interaction cross section appropriate to the interaction of two gamma-rays each of energy 100 MeV: $\sigma_{\gamma\gamma} \approx 10^{-28}$ cm$^2$ (Gould & Schréder 1967). Thus, interactions between gamma-rays produced by inverse Compton scattering are unlikely in the blazar case.

### 2.3. The Formation of Bunches

In the case of injection of positrons in the Klein-Nishina regime (above the line KN-T), the situation is much more complicated, since $\gamma$-rays produced by Compton scattering can interact with soft photons from the funnel giving rise to an electromagnetic cascade. The sequence of events: $\gamma$-ray production by ICS $\rightarrow$ $e^{\pm}$ pair production by the $\gamma$-ray $\rightarrow$ acceleration of the pair to $\gamma_{eq}$ $\rightarrow$ $\gamma$-ray production by ICS, creates an avalanche of pairs similar to that discussed for the case of radio pulsars by Michel (1991). The Lorentz factor $\gamma_{cr}$ corresponding to the boundary between the Thomson and Klein-Nishina regimes can be estimated as

$$\gamma_{cr} \approx \frac{m_e c^2}{(3k_B T)} \approx \frac{2.10^9}{T}. \quad (11)$$

A soft photon of dimensionless energy $x_s = h\nu_s/mc^2$ (with $\nu_s$ the frequency) which is scattered by an electron of Lorentz factor $\gamma_{cr}$ has its energy increased to roughly $x = 4\gamma_{cr}^2 x_s/3$ and is, therefore, just at the threshold for pair production with an unscattered photon, given by $xx_s > 2$. Thus, any pair injected in the Klein-Nishina region will be accelerated to an energy at which the inverse Compton gamma-rays it produces are capable of producing more pairs. In contrast to the mean free path for pair production off another gamma-ray, the mean distance $\lambda_{\gamma\gamma}$ travelled before pair production with a soft photon is very short:

$$\lambda_{\gamma\gamma} \approx (\sigma_{\gamma\gamma} n_{\rm ph})^{-1} \qquad (12)$$
$$\approx \frac{8 \cdot 10^{23}}{T^3}\,{\rm cm}\ , \qquad (13)$$

where $n_{\rm ph}$ is the density of the soft photons and we have approximated the interaction cross section by $\sigma_{\gamma\gamma} \approx 3\sigma_{\rm T}/16$, where $\sigma_{\rm T}$ is the Thomson cross section (as is appropriate for interactions just above threshold). The secondary $e^\pm$ pairs are accelerated by the electric field, positrons moving alongside the seed particle and electrons in the opposite direction. Both types of particle attain a Lorentz factor $\gamma_{\rm eq}$ within a distance $d_{\rm eq}$ given by Eq. (3) and subsequently produce photons capable of creating the next generation of pairs. This cascade continues as long as the particles move in the unmodified external electric field. However, once the accumulation of positive charge around the initial positron is sufficient to screen the external field from the particles immediately behind it, negatively charged particles created in the bunch will no longer be removed. Although the process of pair production by each electron and positron in the bunch continues, increasing the total number of particles present, and thus the total drag force exerted by the soft photons, the net charge of the bunch ceases to rise, reducing the equilibrium value of the Lorentz factor at which the bunch moves. Eventually, the bunch reaches a critical value of the Lorentz factor below which pair creation is no longer possible. Given constant external parameters, we expect the bunch to remain close to this critical Lorentz factor: it cannot move much faster, since then more pairs are created increasing the drag, and it cannot move much slower, since the number of particles would then remain constant, and with it the acceleration and drag forces. This scenario closely resembles that described by Sturrock (1971) and Michel (1991) for the formation of dense almost neutral bunches of $e^\pm$ pair plasma in a pulsar magnetosphere. However, in the case of pulsars, the role of the background thermal radiation is taken over by virtual photons associated with the strong magnetic field.

In order to estimate the total number of pairs in a bunch we follow Michel (1991) and consider a single positron which crosses the KN-T line in the parameter space of Fig. 2. At each step of the cascade, particles are produced which move almost precisely along the electric field direction at almost exactly the speed of light. The spatial dispersion of particles along the field direction is, therefore, extremely small, and the bunch develops into a pancake or disk whose extension perpendicular to the field direction is caused by the small angle ($\sim 1/\gamma_{\rm cr}$) which scattered photons subtend with the direction of the electric field. For each fresh generation, therefore, the new electric field line on which pairs are born is displaced from the old one in a random direction (perpendicular to the field) by a distance

$$\Delta \approx 1/(\sigma_{\rm T} n_{\rm ph} \gamma_{\rm cr}).\ . \qquad (14)$$

After $N$ cascade generations, the random walk across electric field lines produces a disk of radius approximately $\Delta N^{1/2}$, and area $S$ given approximately by

$$S = 1.8 \times 10^{28} N/T^4\ {\rm cm}^2. \qquad (15)$$

The minimum number of positrons $n_+$ required to form a bunch can be estimated assuming all electrons are removed until the field produced by the positron disk exactly balances the externally applied field just behind the disk. Using Gauss' law we find:

$$\begin{aligned} n_+ &= SE^{\rm f}/(4\pi e) \\ &= 10^{34} E^{\rm f} N/T^4. \end{aligned} \qquad (16)$$

The development of the cascade is described in Fig. 3. The number of positrons $n_+(N)$ in the leading front after $N$ interactions is given by the recursion relation

$$n_+(N) = n_+(N-1) + n_+(N-2), \qquad (17)$$

together with the initial values $n_+(0) = 1$, $n_+(1) = 1$. For large $N$ we have approximately

$$n_+(N) \approx \exp[(N-3)/2]. \qquad (18)$$

For a reasonable electric field strength $E^{\rm f} = 4 \times 10^3$ V cm$^{-1}$ and temperature $T = 5 \times 10^5$ K we find, by comparing equations (16) and (18), that saturation sets in at about $N \approx 80$ at which point the front contains some $5 \times 10^{16}$ positrons (see Fig. 4).

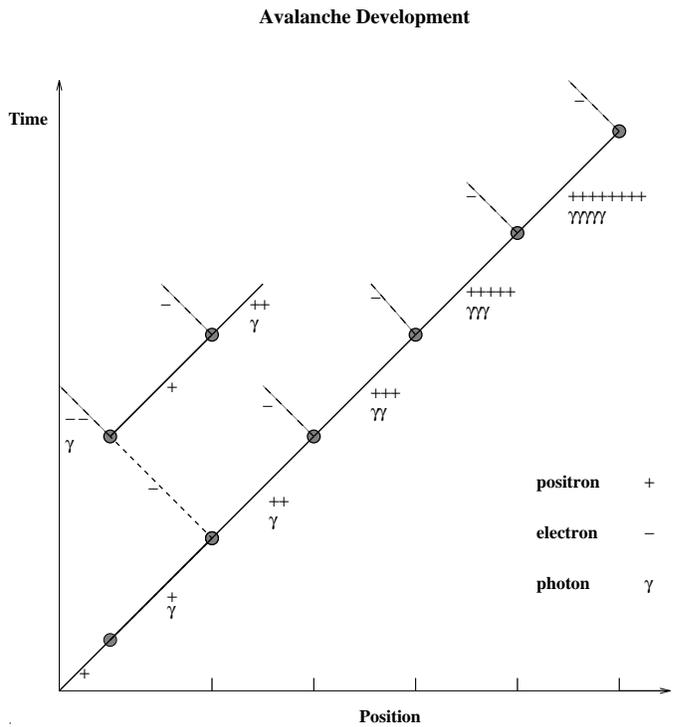

**Fig. 3.** A simplified one-dimensional picture of avalanche development. An interaction (Compton scattering or $\gamma$–$\gamma$ pair production) is assumed to occur after a particle ($e^\pm$ or photon) has travelled exactly one interaction length. The electric field maintains the speed of charged particles at approximately that of light, with electrons moving to the left and positrons to the right

At this point, our simple one dimensional picture has two fronts, one containing $5 \times 10^{16}$ positrons moving outwards and another containing an almost equal number of electrons moving inwards. Between them, there is a fairly smooth distribution of charged particles which, however, find themselves in a region of zero electric field. In reality, however, the finite transverse extent of the disk remains very small compared to an interaction length, as can be seen from Eq. (15), so that the shielding effect of the leading disk will not extend to following disks, but will merely ensure that negatively charged particles subsequently created close to the leading front are no longer removed from it.

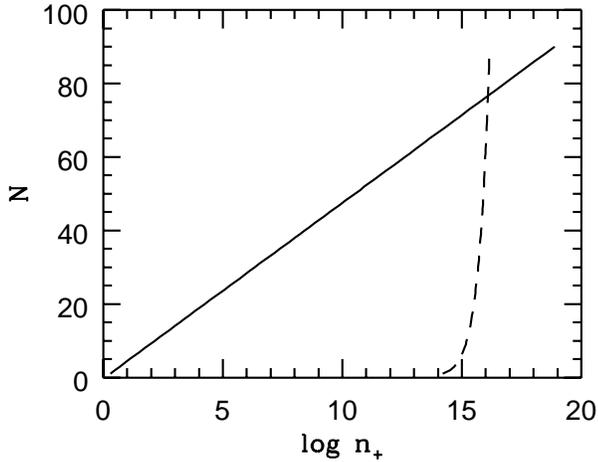

Fig. 4. The almost exponential growth of the number of positrons in a bunch as a function of the number of interactions $N$ (solid line). The dashed line depicts the number of positrons required in the bunch in order to cancel the external electric field after $N$ interactions. The point of intersection gives the estimated bunch dimension. The parameters used are $E^{\rm f} = 4 \times 10^3$ V cm$^{-1}$ and $T = 5 \times 10^5$ K.

Models very similar to this one involving bunches of charged particles have been discussed by several authors in connection with the radiation mechanism of radio pulsars (for a review, see Melrose 1992). There are two main problems with them: (i) whether or not the bunches are stable and (ii) whether or not they are capable of emitting radio waves coherently, as required to explain the radio pulses. Only the first of these points is relevant here. The stability of bunches is, of course closely related to the process responsible for their formation. Early work suggested that the growth of waves in the two-stream or beam-plasma instability could lead to a series of charged bunches. However, these are rather quickly destroyed in the nonlinear saturation phase of the instability. Only in the case of a beam penetrating a completely charge separated background (Elsässer & Kirk 1976) does it seem possible to have bunches which exist for many plasma periods (i.e., many times the inverse of the plasma frequency). Even so, there are difficulties connected with the coherent radiation of such bunches (Asseo et al 1980).

Quite a different approach is required to the problem of bunch stability in the model proposed by Michel (1991). Here it is assumed that regions around a pulsar exist in which a strong electric field oriented parallel to the magnetic field maintains a vacuum gap. In close analogy with the case we describe above, bunches are created out of the vacuum by interaction of a photon with the magnetic field, and the process of curvature radiation by the created pair provides more photons to continue the cascade. The particle energy is limited by radiation reaction, and the number of particles in a bunch exponentiates in time until the oppositely charge lepton starts to be bound to the bunch. Plasma instabilities are not thought to play a major role in the evolution of the cascade, but, instead, work has concentrated on the dynamics of the individual charged particles under the influence of radiation drag and the ambient electric field. Currently, there are two points of view: Fawley (1978), proposes that an electron positron pair injected into such a gap produces two 'flame-fronts' moving in opposite directions along the magnetic field, rather like the outward and inward propagating electron and positron bunches described above. However, on the basis of one-dimensional simulations, he hypothesises that the region between the flame fronts has zero electric field and is filled with an almost neutral plasma. On the other hand, Michel (1991) proposes a picture in which also the interior region between the flame fronts organises itself into bunches. It is this picture which we adopt. The physical justification lies in the inherently two-dimensional nature of the problem: the transverse dimension of a flame front can be estimated (cf. Eq. 15) to be very much smaller that the separation of the fronts. Thus, the trailing cloud of electrons and positrons which are completely shielded from the ambient electric field in the one-dimensional picture must be almost completely exposed to it in reality. It is thus plausible that they also evolve in such a way as to create their own flame fronts, i.e., bunches.

*2.4. The Inverse Compton Emission of Bunches*

Although the stability of the proposed bunches remains to be demonstrated, it is interesting to analyze the fate of a bunch once it has accumulated sufficient charge to bind its electrons. If we define $\gamma_{\rm b}(n_+, n_-)$ to be the equilibrium Lorentz factor of a bunch containing $n_-$ electrons and $n_+$ positrons, we can balance gains and losses to find:

$$\eta e E^{\rm f} = \frac{1}{c} P_{\rm ICS}(\gamma_{\rm b}, T), \qquad (19)$$

where

$$\eta = (n_+ - n_-)/(n_+ + n_-), \qquad (20)$$

$P_{\rm ICS}(\gamma_{\rm b}, T)$ is the energy loss rate a single electron or positron with Lorentz factor $\gamma_{\rm b}$ by inverse Compton scattering of black body photons with temperature $T$, and $-e$ is the charge of an electron. When saturation occurs, the value of the parameter $\eta$ should not be too different from unity: $\eta_0 = \eta(r_{\rm min}) \approx 1$. However, the process of pair production continues to cause a rapid rise in the number of pairs and consequent decrease of $\eta$, so that we expect the bunch to settle to an equilibrium Lorentz factor just inside the Thomson regime: $\gamma_{\rm b} \approx \gamma_{\rm cr}$. We can then use Eq. (19) to determine $\eta$, given that the net charge of the bunch is roughly the number of positrons it contained when electrons started to be bound to it.

The subsequent evolution of the bunch through the funnel is determined by the electric field and temperature profiles. Two cases can be distinguished:

1. If $E^{\rm f} T^{-2}$ increases along the path, $\eta$ decreases steadily and is given by Eq. (19). The Lorentz factor of the bunch is then always equal to $\gamma_{\rm cr}$.
2. If, on the other hand, $E^{\rm f} T^{-2}$ decreases, the Compton drag decelerates the bunch below $\gamma_{\rm cr}$, pair creation stops, and $\eta$ remains constant.

In each case, we can estimate the spectrum radiated by a bunch using the $\delta$–function approximation to Compton scattering, in which the number $F$ of photons of frequency $\nu$ emitted by a single particle per second per unit frequency interval is given by:

$$\frac{\mathrm{d}F}{\mathrm{d}\nu\mathrm{d}t} \approx \frac{4\sigma_\mathrm{T} c \gamma^2 U_\mathrm{rad}}{3h\nu}\delta(\nu - \nu_0), \qquad (21)$$

where $U_\mathrm{rad}$ is the energy density in the background photons and $\nu_0 = 4(3k_\mathrm{B}T)\gamma^2/(3h)$ is the frequency of scattered radiation. Once created, a gamma-ray photon leaves a bunch freely, since the optical depth to scattering through the bunch, $n_+\sigma_\mathrm{T}/S$, is very small (typically $\sim 10^{-15}$), as can be seen from Eqs. (15) and (18). The total emission per unit solid angle $\mathrm{d}\mathcal{F}/\mathrm{d}\nu\mathrm{d}\Omega\mathrm{d}t$ is found by integrating over the path of a bunch:

$$\frac{\mathrm{d}\mathcal{F}}{\mathrm{d}\nu\mathrm{d}\Omega\mathrm{d}t} = \frac{\dot{N}}{c}\int \mathrm{d}r\, N_\mathrm{b}\frac{\mathrm{d}F}{\mathrm{d}\nu\mathrm{d}t} \qquad (22)$$

where $\dot{N}$ is the rate of injection of bunches per unit solid angle, $N_\mathrm{b} = N_0\eta_0/\eta$ is the total number of electrons and positrons in the bunch, $\eta_0 N_0 = (n_+ - n_-)$ is the net bunch charge (which remains constant after bunch formation) and $N_0 = N_\mathrm{b}$ at the moment of bunch formation. The emission of a single particle is beamed into a very small angle ($\approx 1/\gamma_\mathrm{eq}$) about the direction of motion. However, even a slight flaring of the funnel walls should cause the electric field lines and hence the direction of motion of the particles to diverge, allowing an observer to sample the entire cone of single particle emission. Consequently, it is appropriate to use the expression for the emissivity integrated over all emission angles in Eq. (22). Adopting power law profiles according to:

$$\begin{aligned} E^\mathrm{f} &= E^\mathrm{f}_\mathrm{max}\left(\frac{r}{r_\mathrm{min}}\right)^{-p} \\ T &= T_\mathrm{max}\left(\frac{r}{r_\mathrm{min}}\right)^{-q} \end{aligned} \qquad (23)$$

we find the spectra

$$\frac{\mathrm{d}\mathcal{F}}{\mathrm{d}\nu\mathrm{d}\Omega\mathrm{d}t} = \begin{cases} 2.4\times 10^{14}(3.3\times 10^{29})^{(p-1)/q}|q|^{-1} \\ \quad \times \dot{N}N_0\eta_0 r_\mathrm{max}E_\mathrm{max} \\ \quad \times T_\mathrm{max}^{-(p-1)/q}\nu^{-2-(p-1)/q} \\ \quad \text{in case 1 } (p < 2q) \\ \\ 2.4\times 10^{14}(1.9\times 10^{37})^{(p-1)/(3q-p)}|3q-p|^{-1} \\ \quad \times \dot{N}N_0 r_\mathrm{max}(\eta_0 E_\mathrm{max})^{1+(p-1)/(3q-p)} \\ \quad \times T_\mathrm{max}^{-3(p-1)/(3q-p)}\nu^{-2-(p-1)/(3q-p)} \\ \quad \text{in case 2 } (p > 2q) \end{cases} \qquad (24)$$

It is interesting to note that flat spectra (of index $< 2$) can be found for $p < 1$ in case 1, provided the temperature decreases outwards ($q > 0$). In case 2 flat spectra are also obtained for $p < 1$, provided the temperature gradient is sufficiently steep ($q > p/3$).

## 3. Application to $\gamma$-ray emission from blazars

In order to investigate the applicability of a bunch scenario to $\gamma$-ray emitting AGNs, we have fitted the $\gamma$-ray spectrum of two blazars (CTA 102 and QSO 1633+382), recently observed by EGRET (GRO) at energies above 100 MeV. CTA 102 is a $\gamma$-ray emitting AGN with one of the softest spectra so far observed by EGRET (spectral index $2.6\pm 0.2$, Nolan et al. 1993). In contrast, QSO 1633+382 shows one of the hardest spectra in the EGRET sample (spectral index $1.86\pm 0.07$, Mattox et al. 1993). The strategy we adopt to obtain a fit is first to fix the maximum and minimum radii and temperatures, according to ideas concerning the structure of thick accretion disks and to observational indications of the likely temperature range. The parameter $q$ is thereby determined (see Table 1). We then have the two choices indicated in Eq. (24) for obtaining a value of $p$ which gives a good fit to the spectrum. In case 2, in which a bunch moves with Lorentz factor less than $\gamma_\mathrm{cr}$ (defined in Eq. 11) CTA 102 would require $p = 1.8$, which is in contradiction with the condition $p > 2q$, needed for bunch emission. However, this case applies not only to bunches of fixed charge, but also to individual particles, provided they scatter in the Thomson regime. In fact, the required spectrum can be obtained if the particles remain beneath the KN-T line in Fig. 2 and do not form bunches i.e., if the electric field within the funnel is sufficiently small ($E_\mathrm{max} < 7\times 10^3$ V cm$^{-1}$ for the assumed range of temperature). If, on the other hand, the parameters of the funnel imply that particles cross the KN-T, then they immediately form bunches, and, since $p < 2q$, we must employ the formula appropriate to case 1. Since a bunch contains a large number of particles, that part of the emission produced above the KN-T dominates any contribution from the single particle whilst it moved below the line. The value of $p$ required for case 1 is 1.5, consistent with the requirement $p < 2q$.

The situation for the hard spectrum source QSO 1633+382 is similar. Once again, a straightforward application of the formula pertaining to case 2 results in $p = 0.19$, in conflict with the condition $p > 2q$. A fit is possible for individual particles scattering in the Thomson regime (i.e., below the KN-T line in Fig. 2) only if $E_\mathrm{max} < 1$ V cm$^{-1}$. Consequently, we concentrate here on case 1.

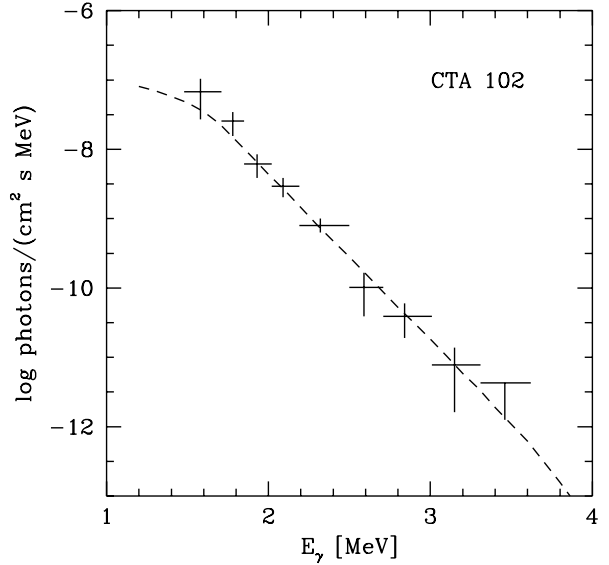

Fig. 5. The $\gamma$-ray spectrum of the blazar CTA 102 ($z \approx 1$.) observed by EGRET (Nolan et al. 1993) and the fit obtained in the bunch scenario, with the parameters given in Table 1.

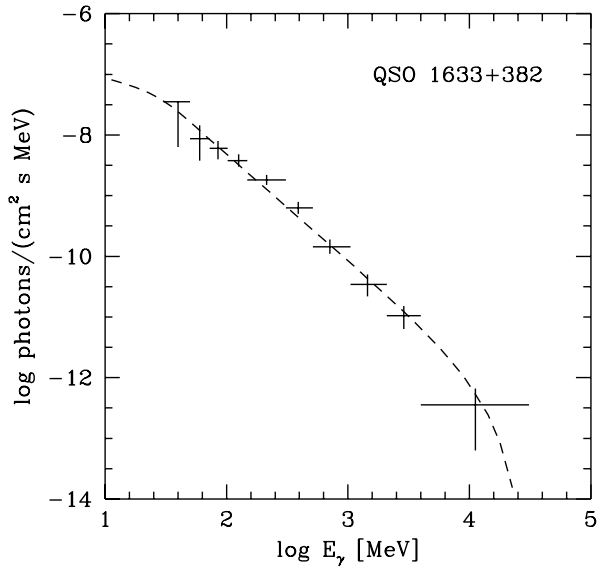

**Fig. 6.** As in Fig. 5, but for the blazar QSO 1633+382 ($z \approx 1.8$) (Mattox et al. 1993)

In calculating the spectrum we relax the delta function approximation to Compton scattering in favour of the formula given by Blumenthal and Gould (1970; Eq. 2.48). The fits to the $\gamma$-ray spectra of CTA 102 and QSO 1633+382 are presented in Figs. 5 and 6. They are obtained with the temperature and electric field profiles given by Eq. (23), using the parameters shown in Table 1. These profiles are also marked in Fig. 2; they correspond to emission by bunches of charge. The fits obtained for emission by single particles in the Thomson regime are not shown. They differ only in that the power-law extends to lower frequencies without the flattening visible in Figs. 5 and 6.

For the models listed in Table 1, we estimate, using Eqs. (16) and (18) and the values of $T_{max}$ and $E^f_{min}$ the number of positrons initially in the bunch to be roughly $10^{15}$ for each source. Then, normalising the model flux to that observed by EGRET, we use Eq. (22) to find the rate of injection of bunches:

$$\dot{N} \approx \begin{cases} 3.10^{22} \text{ bunches sr}^{-1}\text{ s}^{-1} & \text{for QSO1633 + 382} \\ 3.10^{21} \text{ bunches sr}^{-1}\text{ s}^{-1} & \text{for CTA102} \end{cases} \quad (25)$$

where we have assumed the cosmological parameters $\Omega = 0.5$ and $H_0 = 75$ km/s/Mpc. Assuming an opening angle of $5°$ for the funnel, Eq. (25) implies a total of $2.10^{20}$ and $2.10^{19}$ bunches per second for QSO1633+382 and CTA102, respectively.

Figure 7, shows the evolution of the Lorentz factor and particle content of a bunch in the profiles used for modelling the two blazars. For each source, the evolution follows case 1, resulting in a steady increase in the Lorentz factor outwards. The number of particles in the bunch $N_b$ increases too, keeping the scattering just on the boundary between the Klein-Nishina and Thomson regimes.

It is important to note that the model we discuss is able to fit reasonably the broad range of spectral indices in $\gamma$-ray energy range reported from AGNs by EGRET. In general, harder spectra are obtained when a large part of the funnel has a relatively strong electric field and low temperature. The low frequency inverse Compton emission arises in the inner parts of the funnel. Since the emitting particles are accelerated essentially instantaneously to $\gamma_{eq}$, and because they leave the funnel and its field of soft photons without being allowed to cool, there is a lower cut-off in the energy spectrum of radiating particles, which leads to a turn-over in the photon spectrum at about 10 MeV. The inverse Compton spectrum at lower frequencies eventually becomes completely flat (Blumenthal & Gould 1970). However, we do not attempt to explain the spectrum in the hard X-ray region within our model. All the $\gamma$-ray spectra in our model show a cut-off at few tens of GeV.

**Table 1.** Model parameters

|  | CTA 102 | QSO 1633+382 |
|---|---|---|
| $p \; (= -d\ln E^f/d\ln r)$ | 1.5 | 0.65 |
| $q \; (= -d\ln T/d\ln r)$ | 1.27 | 1.42 |
| $T_{max}$ (K) | $3.5 \times 10^6$ | $3.5 \times 10^6$ |
| $T_{min}$ (K) | $10^4$ | $5 \times 10^3$ |
| $E^f_{max}$ (V cm$^{-1}$) | $2 \times 10^5$ | $2 \times 10^5$ |
| $E^f_{min}$ (V cm$^{-1}$) | $2 \times 10^2$ | $10^4$ |
| $r_{min}$ (cm) | $10^{14}$ | $10^{14}$ |
| $r_{max}$ (cm) | $10^{16}$ | $10^{16}$ |

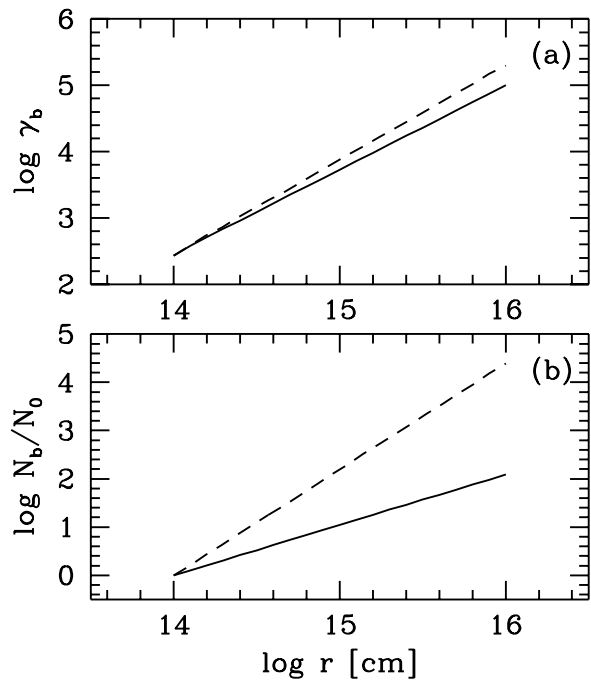

**Fig. 7.** (a) The Lorentz factor $\gamma_b$ and (b) the number of electrons and positrons in the bunch $N_b$ as a function of position in the funnel for the parameters of CTA 102 (solid line) and QSO 1633+382 (dashed line)

This is determined by particles which emit close to the Klein-Nishina/Thomson transition in the outer parts of the funnel, where the temperature is of the order of $10^4$ K. In one case, Mkn 421, $\gamma$-ray emission has been reported at TeV energies (Punch et al. 1992). This source is, however, not typical of the EGRET sample (it is one of the closest objects detected so far by EGRET and has a significantly lower luminosity than observed from other AGNs), and may operate by a different radiation mechanism. In a model such as ours, TeV photons do not arise naturally, since this would require either a much lower temperature in the radiation field or a strong anisotropy (Burns & Lovelace 1982). One possibility – that injected protons are converted into neutrons which decay at large distance from the core – can be discounted: this requires proton injection above the solid line in Fig. 2 in order to convert the charged particle into a neutron. However, the associated positron from pion and muon decay is injected in a region above the KN-T line, and so will produce a bunch containing a large number of positrons (see Eq. (16)). The energy extracted from the electric field by these positrons would by far exceed that taken off by the neutron.

## 4. Summary and conclusions

We investigate the possibility that the $\gamma$-ray emission recently detected from more than 20 AGNs is produced in the narrow funnel of a thick accretion disk by relativistic electrons and positrons which comptonise soft thermal photons from the funnel walls. These particles, injected directly or as secondaries from hadronic interactions, are assumed to be accelerated in a longitudinal electric field which may be induced in the inner parts of the funnel along magnetic field lines which thread the horizon of a rotating supermassive black hole. Other scenarios which produce aligned magnetic and electric electric fields (such as reconnection in an MHD jet) produce basically similar effects, but the details of the spectrum may be different. As a result of the balance between energy losses by inverse Compton scattering and energy gains from electric field, electrons and positrons move through the funnel in either the Thomson or Klein-Nishina regimes of scattering (see Fig. 2 for the corresponding parameter space in $T$ and $E^f$ coordinates). In the Thomson regime, the electrons move with the equilibrium Lorentz factor defined in Eq. (2) and efficiently comptonise soft photons into the $\gamma$-ray energy range.

The propagation of electrons in the Klein-Nishina regime is much more complicated. We argue that single electrons produce dense bunches of $e^\pm$ plasma containing relatively low net electric charge. A bunch, on its subsequent path through the funnel, may grow by producing captive pairs or decelerate into the Thomson regime, according to the $T$ and $E^f$ profiles in the funnel. Our bunch production scenario resembles the production of bunches of $e^\pm$ plasma in pulsar magnetospheres proposed by Michel (1991).

A bunch moving through the funnel extracts energy from the electric field and converts it via inverse Compton scattering into $\gamma$-ray photons. The Lorentz factor of the bunch is self-regulated such that it is either close to or below the value at which the scattered photons could produce pairs on interacting with the thermal photons. As a result, softer $\gamma$-rays are produced in regions where $T$ is higher (in the inner part of the funnel) and harder $\gamma$-rays in regions of lower T (the outer parts of the funnel). We model the $\gamma$-ray emission from blazars in a bunch scenario (sect. 3), fitting, as examples, the $\gamma$-ray spectra of two blazars (CTA 102, Fig. 5 and QSO 1633+382, Fig. 6). However, it is also possible to obtain fits to the spectra without forming bunches. This requires radiation by single particles which scatter the soft photons in the Thomson regime. In order that pair creation be avoided, a relatively weak electric field is necessary (see section 3) – for the hard spectrum source QSO 1633+382 this condition is particularly restrictive. In our model, the spectrum below the $\gamma$-ray region flattens, so that we do not produce the nonthermal X-ray and optical emission. If we were to increase the radiation temperature in the inner parts of the funnel, X-ray photons would result, but this would imply a thermal component of much higher temperature than usually observed in the soft X-ray excesses of AGNs. Instead, we propose that the nonthermal X-rays and optical photons are emitted outside the funnel either by comptonisation of the anisotropic radiation from the funnel, or by synchrotron radiation. A basic problem encountered by all models of gamma-ray emission from blazars is that of absorption by the hard X-ray component. Since our source of gamma-rays is close to the central object, an isotropic distribution of hard X-ray photons would absorb it completely. This problem can be avoided only by assuming the nonthermal X-rays are beamed in the same direction as the gamma-rays. A beaming angle of 5°, for example, implies that a 100 MeV gamma-ray requires a target photon of 1 MeV in order to pair produce, whereas a 5 keV target suffices in the isotropic case. From Eqs. (7) and (9), one then finds that the effect of absorption is negligible.

If protons are injected deep in the funnel into the acceleration region, their interaction with soft photons is by direct production of $e^\pm$ pairs and/or $\pi$'s (which then decay, producing $e^\pm$ pairs). The relative importance of these two processes depends on the $T$ and $E^f$ in the injection region (these two regions are separated in Fig. 2 by the line $p\gamma \to \pi/e^\pm$). There are significant differences between direct injection of protons and electrons in the acceleration region:

1. The $e^\pm$ pairs, produced as a secondaries by the decay of $\pi$'s, have such high Lorentz factors that the process of triplet $e^\pm$ pair production determines the bunch formation. However, the fate of the bunch is similar to the case of direct injection of low energy electrons.
2. The interaction of protons in the regime of direct $e^\pm$ pair production (below the line $p\gamma \to \pi/e^\pm$) produces an injection of pairs along the whole path of proton's motion in the funnel. If this happens below the line $E_\gamma(> / < 0.5 \text{MeV})$, the $e^\pm$ pairs are slowed down to the Thomson regime and comptonise the soft thermal photons into the hard X-ray energy range. If the injection of pairs occurs above this line, the $\gamma$-ray photons, produced in the comptonisation process, can, in principle, interact with each other to produce pairs (electrons and positrons produce beams of $\gamma$-rays moving in opposite directions). The timescale of this instability depends on the optical depth in the funnel. If it is short enough, the entire funnel is filled with dense low energy $e^\pm$ plasma which may partially annihilate (producing broad annihilation feature) and may be partially expelled outside the funnel by radiation pressure (a mildly relativistic $e^\pm$ jet may be created). In blazars, this optical depth is very small, but the scenario could conceivably be applied to the recently reported transient source of annihilation radiation in the Galactic Center 1E1740.7-2942 (Bouchet et al. 1991;

Sunyaev et al. 1991). However, we do not pursue this line of investigation here.

3. Almost all of the protons, which interact with soft photons in the region characterised by parameters $T$ and $E^f$ above the line $p\gamma \to \pi/e^{\pm}$, are converted after a few interactions into relativistic neutrons. The neutrons can leave the funnel without significant energy losses because they are below the threshold for $\pi$ production in the outer regions of the funnel ($T$ and $E^f$ decreases towards the outer parts of the funnel). They decay at a distance determined by their Lorentz factor. However, a much larger luminosity is expected of the bunch produced by the positron resulting from decay of the associated $\pi^+$.

Since plasma escaping from the funnel carries a net electric charge, it may be supposed that the central black hole (and/or inner parts of the accretion disk) will accumulate an opposite charge after certain time, preventing the stationary operation of the picture discussed here. However, as noted by Lovelace (1976) this charge can be neutralised by a flow of current through the accretion disk.

As showed in Figs. 5 and 6, our model can provide high energy $\gamma$-ray spectra with a the broad range of spectral indices. However, the spectra show a characteristic sharp cut-off at an energy of a few tens of GeV. This is determined by the temperature in the outer parts of the funnel. It is difficult to determine at present, on the basis of CGRO observations, if such a cut-off is indeed present in the spectra of AGNs. One clear counter-example is Mrk 421 which has been detected at TeV energies. However, this source is not a typical member of the sample of powerful $\gamma$-ray emitting AGNs reported by EGRET.

*Acknowledgements.* W.B. thanks the Max-Planck-Institut für Kernphysik for hospitality during his stay in Heidelberg and the Deutsche Forschungsgemeinschaft for support under Sonderforschungsbereich 328. We are grateful to A. Mastichiadis for helpful discussions and to K. Mannheim for detailed and constructive suggestions.